\documentclass[aps,prb,twocolumn,preprintnumbers,superscriptaddress,eqsecnum,showpacs,amsmath,amssymb]{revtex4}
\usepackage{graphicx}
\usepackage{bm}
\usepackage{natbib}

\usepackage[breaklinks,colorlinks,bookmarks=false,citecolor=blue,linkcolor=red,urlcolor=blue]{hyperref}

\newcommand{\beqa}{\begin{eqnarray}}
\newcommand{\eneqa}{\end{eqnarray}}
\newcommand{\beq}{\begin{equation}}
\newcommand{\eneq}{\end{equation}}
\newcommand{\nn}{\nonumber}
\newcommand{\ua}{\uparrow}
\newcommand{\da}{\downarrow}

\begin{document}

\preprint{\href{https://doi.org/10.1103/PhysRevB.81.115416}{Phys. Rev. B {\bfseries 81}, 115416 (2010)};
\href{https://doi.org/10.1103/PhysRevB.101.049909}{Phys. Rev. B {\bfseries 101}, 049909(E) (2020)}
}

\title{Magnetism of Finite Graphene Samples: Mean-Field Theory compared with Exact
    Diagonalization and Quantum Monte Carlo Simulation}
\author{H\'el\`ene Feldner}
\affiliation{Institut de Physique et Chimie des Mat\'eriaux de Strasbourg, UMR7504, CNRS-UdS,
23 rue du Loess, BP43, 67034 Strasbourg Cedex 2, France}
\author{Zi Yang Meng}
\affiliation{Institut f\"{u}r Theoretische Physik, Universit\"{a}t Stuttgart, 70550 Stuttgart, Germany}
\author{Andreas Honecker}
\affiliation{Institut f\"{u}r Theoretische Physik,
          Georg-August-Universit\"{a}t G\"{o}ttingen,
          Friedrich-Hund-Platz 1, 37077 G\"{o}ttingen, Germany}
\author{Daniel Cabra}
\affiliation{Institut de Physique et Chimie des Mat\'eriaux de Strasbourg, UMR7504, CNRS-UdS,
23 rue du Loess, BP43, 67034 Strasbourg Cedex 2, France}
\author{Stefan Wessel}
\affiliation{Institut f\"{u}r Theoretische Physik, Universit\"{a}t Stuttgart, 70550 Stuttgart, Germany}
\author{Fakher F. Assaad}
\affiliation{Institut f\"{u}r Theoretische Physik und Astrophysik,
 Universit\"{a}t W\"{u}rzburg, Am Hubland, 97074 W\"{u}rzburg, Germany }

\date{October 28, 2009; revised January 15, 2010; corrected January 5, 2020}

\begin{abstract}
The magnetic properties of graphene on finite geometries are studied using
a self-consistent mean-field theory of the Hubbard model. This approach is
known to predict ferromagnetic edge states close to the zig-zag edges
in single-layer graphene quantum dots and nanoribbons. In order to
assess the accuracy of this method, we perform complementary
exact diagonalization and quantum Monte Carlo simulations. We observe
good quantitative agreement for all quantities investigated provided that
the Coulomb interaction is not too strong.

\end{abstract}

\maketitle

\section{Introduction}

Graphene consists of a single layer of carbon atoms arranged in
a honeycomb crystal lattice \cite{graphene} and is a promising material
with unique electronic properties. Among the
most important characteristics, one should mention the presence of massless
carriers, weak spin-orbit coupling, insensitivity to an
external electrostatic potential (Klein paradox), fractional quantum Hall
effect, etc.\ (for a review of the main features of graphene see
Ref.~\onlinecite{revue}). 
The electronic properties of graphene nanostructures such as nanoribbons or quantum dots are expected to be
very different from bulk graphene. In fact, the Coulomb interaction
is considerably enhanced in smaller geometries such as quantum dots,
leading for example to unusual blockade effects.\cite{Coulomb-enhanced-3,Coulomb-enhanced-2,Coulomb-enhanced-1}
On the other hand, the edge effect, which depends strongly on the geometry of the sample boundary,
modifies the electronic structure of graphene.\cite{ES-ribbon-1,Electronic-Edge-2,Electronic-Edge-1}
In particular, it has been predicted that finite graphene
samples can exhibit magnetic edge states
(see, e.g., Refs.\
\onlinecite{ES-ribbon,ES-ribbon-X,DMRG,ab-initio,ES-ribbon-2,magnano,ESBS,ES-ribbon-3,ES-ribbon-4,invRibbon,strained-dot})
suggesting potential spintronics applications of graphene
nanodevices.\cite{spintronics}

It is common practice to use a mean-field theory (MFT) of the
Hubbard model to investigate the magnetic properties of graphene in finite
geometries (see, e.g.,
Refs.~\onlinecite{ES-ribbon,ES-ribbon-X,ES-ribbon-2,magnano,ESBS,ES-ribbon-3,ES-ribbon-4,invRibbon,strained-dot}).
Such a MFT is applicable to any interaction
and any geometry in a quite economic way: within the self-consistent
mean-field (MF) approximation the main numerical effort is to solve the
{\em single-electron} problem on a finite lattice.
However, as far as we are aware, very little is known about the
accuracy of this approximation.

The main purpose of the present paper is to address this issue and check the
accuracy of the MFT. 
We start by recalling a real-space formulation of the MFT
in Sec.~\ref{sec:Model}. In Sec.~\ref{sec:Results}
we briefly look at periodic boundary conditions\cite{MHT}
and show that we can reproduce edge-ferromagnetism for a dot with
zig-zag edges.\cite{magnano,ESBS,strained-dot}
The accuracy of the MFT is carefully examined in Sec.~\ref{sec:Accuracy}
where we present a comparison
with exact diagonalization for a small ``dot'' and
quantum Monte Carlo (QMC) simulations on a larger system with periodic
boundary conditions. We conclude with a summary and perspectives in
Sec.~\ref{sec:Conclusion}.



\section{Model and Computation}

\label{sec:Model}

Since we are interested in the magnetic properties of graphene,
interactions should be taken into account. To this end, we study
the Hubbard model whose Hamiltonian reads
\beq
H=-t\sum_{\langle i,j\rangle ,\sigma}c^{\dagger}_{i,\sigma}c_{j,\sigma}+U\sum_i n_{i,\uparrow}n_{i,\downarrow}
\label{Hubbard}
\eneq
with $n_{i,\sigma}=c^{\dagger}_{i,\sigma}c_{i,\sigma}$.
$\langle i,j\rangle $ are nearest neighbors  on a honeycomb lattice.
We denote the total number of sites by $N$ and the number of electrons
with a spin projection $\sigma=\uparrow,\downarrow$ by
$N_\sigma$. 

Due to the exponential growth of the Hilbert space dimension with
$N$, a direct exact diagonalization of the Hubbard model (\ref{Hubbard})
at half filling is only possible for system sizes until about 20 sites.
In order to deal with larger system sizes we use a MF approximation:
\beqa
H^{MF}&=& -t\sum_{\langle i,j\rangle ,\sigma}c^{\dagger}_{i,\sigma}c_{j,\sigma}
\label{HMF}\\
&&
+U\sum_i \left( \langle n_{i,\uparrow} \rangle n_{i,\downarrow} 
 + n_{i,\uparrow}\langle n_{i,\downarrow}\rangle
- \langle n_{i,\uparrow}\rangle \langle n_{i,\da}\rangle \right)\, . 
\nn
\eneqa
It should be noted that the MF approximation breaks the SU(2)-symmetry of
the original Hubbard model (\ref{Hubbard}).

We compute the ground state
\beq
|GS\rangle =\prod_{\alpha \leq N_\uparrow}d^{\dagger}_{\uparrow \alpha}\prod_{\beta\leq N_\downarrow}d^{\dagger}_{\downarrow \beta}|0\rangle 
\, , \quad
d_{\sigma,\alpha} = \sum_i Q^*_{\sigma,\alpha i}c_{\sigma,i}
\label{GS} 
\eneq
and the one-electron spectrum $\epsilon_{\sigma,\alpha}$ of the
MF Hamiltonian (\ref{HMF}) using the LAPACK library.

This yields the ground-state energy, the local density
\beq
\langle n_{\sigma,i}\rangle =\sum_{\alpha\leq N_\sigma}Q^*_{\sigma,\alpha i}Q_{\sigma,i \alpha} \, ,\label{density}
\eneq
the local magnetization $\langle S^z_i \rangle =
\frac{1}{2} \, \langle n_{i,\uparrow} - n_{i,\downarrow} \rangle$,
as well as the spin correlation functions
\begin{eqnarray}
\langle S^z_i S^z_j\rangle
&=&  \frac{1}{4}\biggl(
\sum\limits_{\sigma=\ua,\da}
\sum\limits_{\alpha=1}^{N_\sigma}\sum\limits_{\beta\neq \alpha}^{N_\sigma}Q_{\sigma,j\alpha}Q_{\sigma,i\beta}
\nonumber \\
&& \quad\times\left\{Q^{*}_{\sigma,\alpha j}Q^{*}_{\sigma,\beta i}-Q^{*}_{\sigma,\beta j}Q^{*}_{\sigma,\alpha i}\right\}
\nonumber \\
&& \quad-\langle n_{j\ua}\rangle \langle n_{i\da}\rangle -\langle n_{j\da}\rangle\langle n_{i\ua}\rangle \biggr)
\nonumber \\
\langle S^x_i S^x_j\rangle&=&\langle S^y_i S^y_j\rangle
=-\frac{1}{4}\biggl(\sum\limits_{\alpha=1}^{N_\ua}\sum\limits_{\beta=1}^{N_\da}Q_{\ua,j\alpha}Q_{\da,i\beta}Q^{*}_{\da,\beta j}Q^{*}_{\ua,\alpha i}
\nonumber \\
&&\qquad\qquad\quad
+Q_{\ua,i\alpha}Q_{\da,j\beta}Q^{*}_{\da,\beta i}Q^{*}_{\ua,\alpha j} \biggr)
\label{SSinej}
\end{eqnarray}
for $i\neq j$ and
\begin{eqnarray}
\langle S^z_i S^z_j\rangle&=&\langle S^x_i S^x_j\rangle=\langle S^y_i S^y_j\rangle
\nonumber \\
&=&
\frac{1}{4}\biggl(\langle n_{i\ua}\rangle+\langle n_{i\da}\rangle-2\langle n_{i\ua}\rangle\langle n_{i\da}\rangle\biggr)
\label{SSi=j}
\end{eqnarray}
for $i=j$.

Self-consistency requires that
the expectation values $\langle n_{\sigma,i}\rangle$ entering
(\ref{HMF}) are equal to the expression (\ref{density}) derived from this
Hamiltonian. We solve this condition iteratively using 
suitable initial conditions
with given numbers of electrons $N_\sigma$.
To overcome convergence problems, we use a
thermal state compatible with the Fermi-Dirac 
distribution at a given temperature  instead of the ground state for the first
iterations.\cite{NWDC} In this case the average density is computed as
\beqa
\langle n_{\sigma,i}\rangle =\sum_{\alpha\in\Omega_\sigma}Q^*_{\sigma,\alpha i}Q_{\sigma,i\alpha}
\, ,\nn
\eneqa
where $\Omega_\sigma$ is a set of $N_\sigma$ single-particle states chosen
randomly with probability
$n(\epsilon_{\sigma,\alpha})=1/(1+e^{[\beta(\epsilon_{\sigma,\alpha}-\bar{\epsilon})]})$.


\section{Results of the mean-field approximation}

\label{sec:Results}

\subsection{System with periodic boundary conditions}

First we briefly discuss the MFT for the infinite system with periodic boundary
conditions. If we assume a N\'eel-ordered configuration,
we find a Mott-Hubbard phase transition at the literature
value $U_c \approx 2.23\,t$,\cite{MHT} where the system goes from a
paramagnetic semi-metal to an insulator with antiferromagnetic order.
The asymptotic behavior of the N\'eel order parameter
and the single-particle gap is numerically consistent with
linear behavior in $U-U_c$ for $U>U_c$, {\em i.e.}, associated
critical exponents equal to one. These unusual mean-field exponents
reflect the unusual density of states\cite{MHT} which is linear close
to the Fermi energy (compare also Ref.\ \onlinecite{exponents}).

We use the critical value $U_c$ mainly
to choose an approximate value of $U$ to describe graphene.
In fact, the correct value of the Coulomb interaction in graphene
is not yet known. Taking the
value of $U$ in polyacetylene, where $U=10$eV and $t=2.5$eV,
suggests $U \approx 4\,t$ for graphene.\cite{revue} Since, at
the MF level, this
value locates the system well inside the antiferromagnetic phase and
it is observed that large graphene sheets do not show magnetic
order, we have decided to use a value of $U$ smaller than $U_c$,
$U = 2\,t$ for the following computation.


\begin{figure}[t!]
    \centering
        \includegraphics[width=0.9\columnwidth]{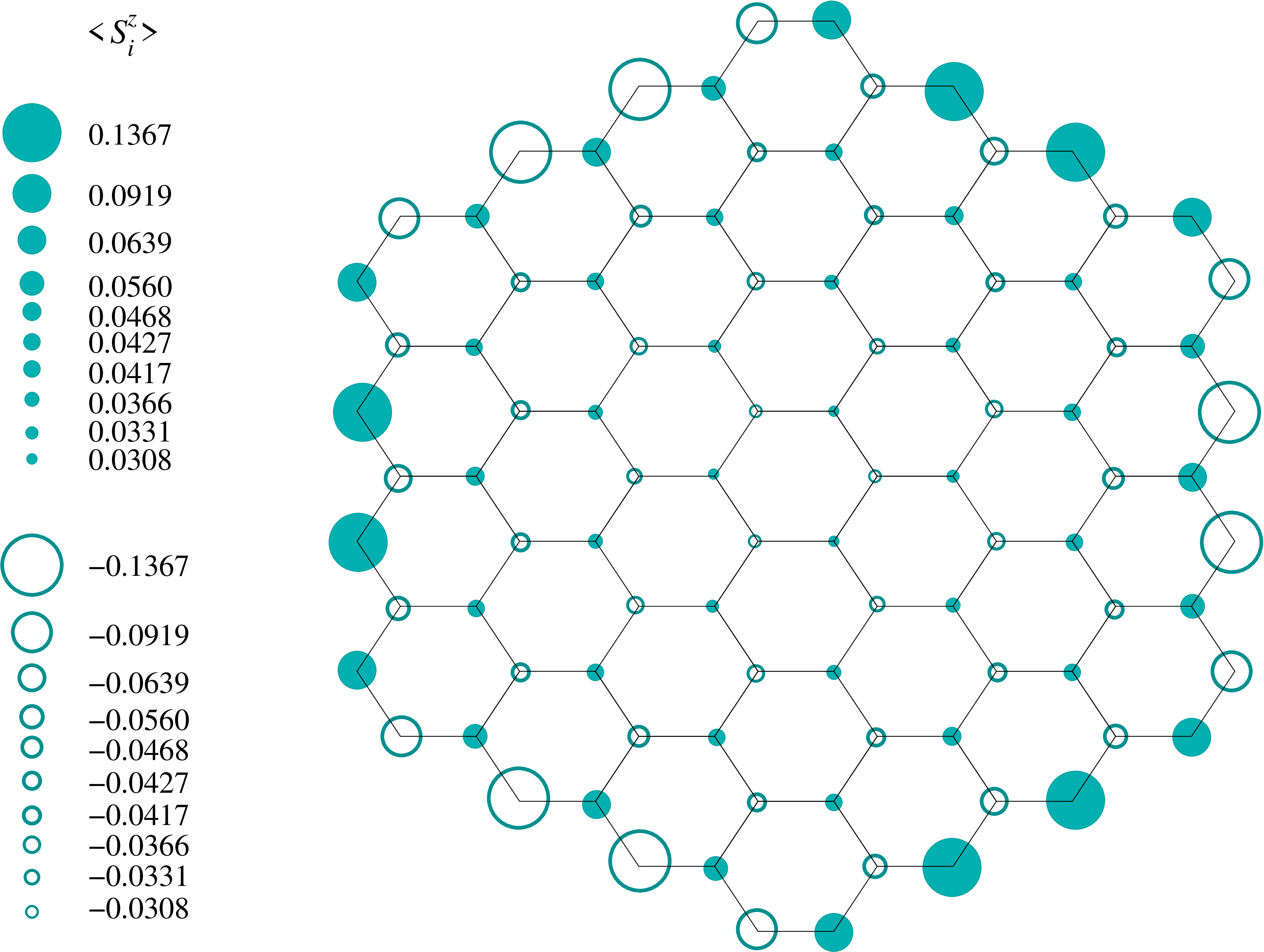}
    \caption{(Color online) Mean-field result for
    the edge magnetization of a hexagonal graphene quantum
     dot with $N=96$ sites and zig-zag edges (at half filling and with $U=2\,t$).}
   \label{fig:edge-ferro-dot}
\end{figure}

\subsection{Edge magnetism on zig-zag edge}

It is well known that even for values of
$U$ smaller than the critical value $U_c$, one
observes a form of ferromagnetism on the 
zig-zag edge of a graphene ribbon\cite{ES-ribbon,ES-ribbon-2,ES-ribbon-3,ES-ribbon-4}
or a quantum dot.\cite{magnano,ESBS,strained-dot}
As an example, Fig.~\ref{fig:edge-ferro-dot} shows our results for the local
magnetization of a hexagonal dot with 96 sites. One observes local ferromagnetic
behavior at each zig-zag edge. By contrast,
systems with armchair edges do not show specific
magnetic properties and follow an evolution closer to the one of a
system with periodic boundary conditions. The difference between the
two edges appears to be a consequence of the fact that in the zig-zag
case only one sublattice is represented on the edge while in
the armchair case both sublattices are present.
Detailed explorations\cite{ESBS,ES-ribbon-4}
demonstrated that the ferromagnetism of zig-zag edges resists to armchair
defects and appears already for short edges.



\section{Accuracy of the approximation}

\label{sec:Accuracy}

\begin{figure}[t!]
    \centering
        \includegraphics[width=1\columnwidth]{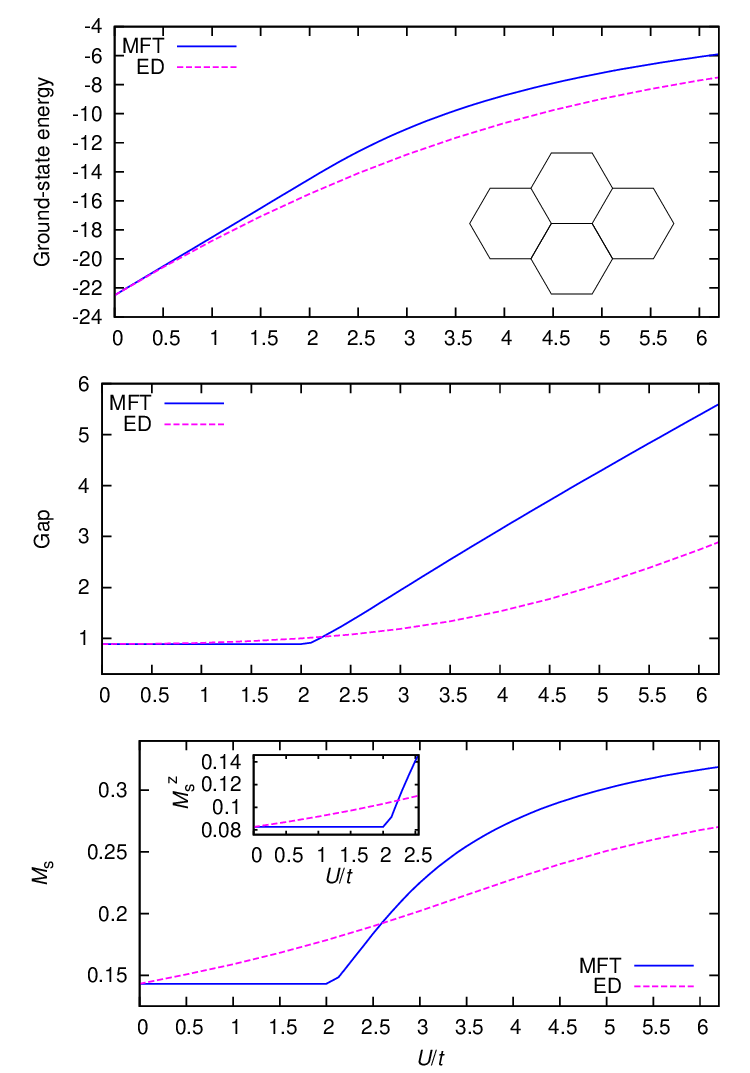}
    \caption{(Color online) Comparison  MFT-ED for the finite-size system of 16 sites sketched
     in the inset of the top panel
     at half filling.
     The bottom panel shows the total staggered magnetization $M_{\rm s}$
     in the main panel and the $z$-component $M_{\rm s}^z$ in the inset.}
    \label{fig:MFT-ED}
\end{figure}

\subsection{Comparison with exact diagonalization for open boundary conditions}

To verify the accuracy of the MFT we first
compare the results with those obtained with exact diagonalization (ED)
of the Hubbard model which was performed with Spinpack.\cite{spinpack}
Due to the exponential growth of the
Hilbert space, ED is limited to very small systems.
Here we have studied the dot-like cluster of 16 sites shown in the inset
of the top panel of Fig.~\ref{fig:MFT-ED}.
The following quantities were computed:
(i) The ground-state energy.
(ii) The charge gap defined as $\Delta E = E_{N-1} - 2 E_{N} + E_{N+1}$,
where $E_n$ is the ground-state energy in the sector with $n$ electrons.
(iii) The $z$- and total staggered magnetization as defined in terms
of the longitudinal and total spin structure factor
\begin{eqnarray}
M_{\rm s}^z &=& \frac{1}{N}\,\sqrt{\sum\limits_{i,j}(-1)^{i-j}\,\langle S^z_i S^z_j\rangle} \, ,
\label{eq:defMsz} \\
M_{\rm s} &=& \frac{1}{N}\,\sqrt{\sum\limits_{i,j}(-1)^{i-j}\,\langle \vec{S}_i \cdot \vec{S}_j\rangle} \, .
\label{eq:defMsVec}
\end{eqnarray}
Here, $(-1)^{i-j}$ is a short-hand notation for $+1$ ($-1$) if $i$ and $j$
belong to the same (different) sublattice(s). Within MFT,
the correlation functions
appearing in (\ref{eq:defMsz}) and (\ref{eq:defMsVec})
are computed from (\ref{SSinej}) and (\ref{SSi=j}). In a numerical solution of
the Hubbard model respecting SU(2) symmetry one finds
$M_{\rm s}^z = M_{\rm s}/\sqrt{3}$.

Figure~\ref{fig:MFT-ED} shows a comparison of
ground-state energy, charge gap, and the two  staggered magnetizations
computed both with MFT and ED at half filling, {\it i.e.}, a total
of 16 electrons. As expected,
the two methods yield identical results for $U=0$ and the
MF ground-state energy is always above the exact answer.
The results for all three quantities
stay close for $U \lesssim 2\,t$. This supports the applicability
of MFT at least as a semi-quantitative method
in particular for the
parameters of the dot shown in Fig.~\ref{fig:edge-ferro-dot}.


\begin{figure}[t!]
    \centering
        \includegraphics[width=1\columnwidth]{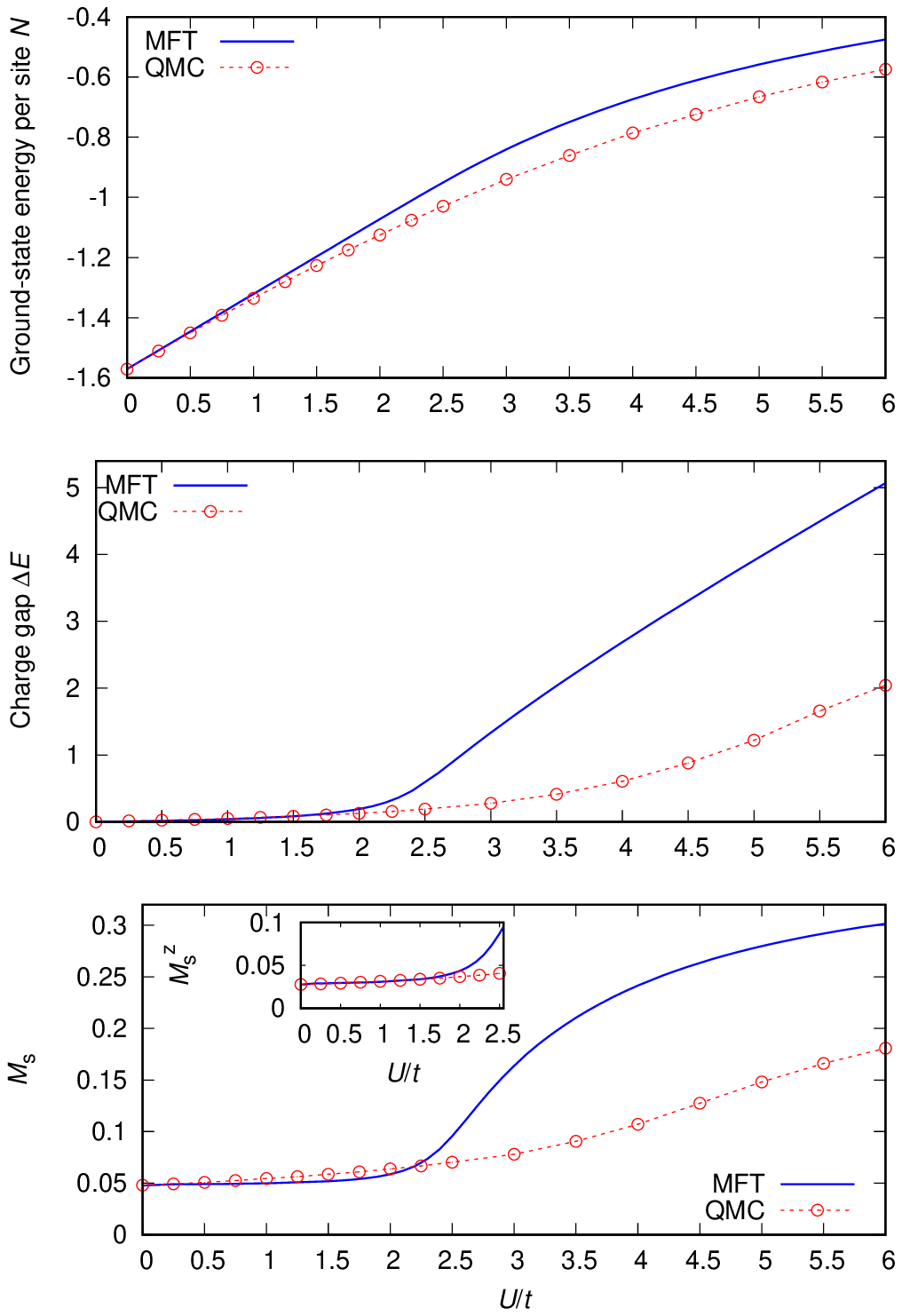}
    \caption{(Color online) Comparison MFT-QMC for a system
    with periodic boundary conditions and $N=162$ sites
    at half filling. QMC error bars are
    smaller than the size of the symbols.}
    \label{fig:MFT-QMC}
\end{figure}

\subsection{Comparison to Quantum Monte Carlo for periodic boundary conditions}

In order to assess the quality of the MFT for larger but still
finite systems, we employ QMC simulations.
We use a projective determinantal QMC approach\cite{assaad08} to obtain
ground-state properties at half filling.
Within this scheme,
expectation values of a physical observable $A$ are obtained from
\beq
{\langle  A \rangle }
=\lim_{\Theta\rightarrow \infty}\frac{{\langle \Psi_T | e^{-\Theta H/2} A
e^{-\Theta H/2} | \Psi_T\rangle }}
{{\langle \Psi_T|e^{-\Theta H}  | \Psi_T\rangle}} \, ,
\label{A}
\eneq
where the trial wavefunction $|\Psi_T \rangle$ must be non-orthogonal to the ground state and
$\Theta$ corresponds to a projection parameter.
We found $\Theta = 40/t$ to be sufficient to obtain converged ground-state quantities within the statistical uncertainty.
For the presented simulations, $\Theta$ was split into discrete step $\Delta\tau$ in the Trotter decomposition.
We verified by extrapolating $\Delta\tau\to 0$ that taking $\Delta\tau=0.05/t$ produced no discretization artifacts.
The simulations were performed on systems with periodic boundary conditions.

Figure~\ref{fig:MFT-QMC} shows our QMC and MFT data for a
finite system with $N=162$ lattice sites.
The top panel of Fig.~\ref{fig:MFT-QMC} shows the energy per site.
In the middle panel of Fig.~\ref{fig:MFT-QMC}
we compare the charge gap $\Delta E$ between
MFT and QMC. In the QMC simulations the single-particle gap
$\Delta_{\rm sp}(\vec{k})$ was obtained by fitting the exponential tail of the imaginary-time displaced Green's function
$G(\vec{k},\tau)\propto \exp(-\tau\Delta_{\rm sp}(\vec{k}))$ at
large imaginary time $\tau$. The single-particle gap
$\Delta_{\rm sp}$ shown in Fig.~\ref{fig:MFT-QMC} equals
$\Delta_{\rm sp}(K)$, {\it i.e.}, the smallest excitation gap
at the Dirac points. Since the system
under study is a particle-hole symmetric half-filled
system, the chemical potential is right in the middle of
the charge gap such that the charge gap is in
fact twice the single-particle gap
$\Delta E = 2 \, \Delta_{\rm sp}$.
Finally, the bottom panel of Fig.~\ref{fig:MFT-QMC} compares the QMC results for the
the total staggered magnetization
Eq.~(\ref{eq:defMsVec}) (main panel)
and the $z$-component Eq.~(\ref{eq:defMsz}) (inset)
with the MFT results.

Again, the MFT follows the QMC results closely for $U \lesssim 2\,t$.
In the present case, the MF curves for the energy, charge gap
$\Delta E$, and $M_{\rm s}^z$
are always above the QMC curves. In fact, one observes
that in the regime $U \lesssim 2\,t$
the agreement between MFT and QMC is a bit better for $M_{\rm s}^z$
than for $M_{\rm s}$. This can be attributed to the MF approximation
explicitly breaking the SU(2) symmetry.
Indeed, in this case one finds that the MF
contribution of the $x$- and $y$-components to $M_{\rm s}$
are independent of $U$ for $U>0$ whereas the $z$-component
increases with increasing $U$.

Appreciable quantitative differences can be observed
in Fig.~\ref{fig:MFT-QMC} at
large $U$ in particular in $\Delta E$ and $M_{\rm s}$.
Indeed, MFT is known to underestimate the stability-range of the
paramagnetic semi-metal by about a factor 2,\cite{MHT,Furukawa01}
{\it i.e.}, quantitative differences are expected for $U$
larger than the mean-field critical value $U_c \approx 2.23\,t$.
Furthermore, in the limit $U \to \infty$ we expect to recover
the $S=1/2$ Heisenberg model where it is known
(compare, e.g., Refs.\ \onlinecite{Heisenberg2D,HeisenbergHoneycomb})
that a full quantum mechanical treatment of the quantity
defined in Eq.~(\ref{eq:defMsVec}) yields a value which for large $N$ is
only about 55\%\ of the classical ({\it i.e.}, MF) value $M_{\rm s}=1/2$.
The difference at the right boundary of the bottom panel of
Fig.~\ref{fig:MFT-QMC} is indeed of this order.
While the MFT and QMC deviate in the precise position of the
quantum critical point,  they agree on locating the system in the paramagnetic
semi-metallic phase for $U \le U_c \approx 2.23\,t$ and the
correspondence 
is at least semi-quantitative for a finite-size system and
$U \lesssim 2\, t$.



\section{Conclusion and perspectives}

\label{sec:Conclusion}

We investigated a self-consistent mean-field approximation
to the Hubbard model on the honeycomb lattice, concentrating on
half filling. The infinite system exhibits a Mott-Hubbard
transition from a paramagnetic semi-metal for a Coulomb repulsion
$U<U_c$ to an antiferromagnetic
insulator for $U > U_c$ with a MF critical value $U_c \approx 2.23\,t$.\cite{MHT}
The mean-field critical exponents associated
to the gap and N\'eel order parameter are numerically consistent with
the value one.\cite{MHT}

We studied the accuracy of the MFT for finite-size systems
with complementary exact diagonalization and
quantum Monte Carlo simulations of the Hubbard model. We  computed the
ground-state energy, the single-particle gap, and the staggered magnetizations
obtained from the spin correlation functions. The MFT 
reproduces the qualitative behavior found
by the other two methods. Furthermore, the quantitative agreement is
reasonable for $U \lesssim 2\,t$, {\it i.e.}, the region which is identified
as a paramagnetic semi-metal both by MFT and QMC. For large values of
$U$, quantitative differences become appreciable. However, the
latter regime corresponds to an antiferromagnetic insulator,
not relevant to graphene.

A weak-coupling instability to a canted antiferromagnet emerges in graphene
when an in-plane  magnetic field  is turned on.\cite{bercx09}
This weak-coupling instability can be captured at the mean-field level
and is confirmed by  QMC simulations.\cite{bercx09} Being equally a weak-coupling
phenomenon, we believe that it will be possible to observe
edge ferromagnetism in future QMC
simulations with zig-zag boundary conditions.
Nevertheless, the MFT has access
to bigger systems than QMC since it is numerically less demanding.
Here we have presented the first piece of evidence that the MFT may be
expected to be quantitatively reliable for $U \lesssim 2\,t$.

\begin{acknowledgments}
We wish to thank T.~C.\ Lang for fruitful discussions,
NIC J\"ulich and HLRS Stuttgart for the allocation of computer time,
and the ESF for financial support through the INSTANS program.
A.H.\ acknowledges support by the Deutsche Forschungsgemeinschaft through
grant HO~2325/4-1 and S.W.\ and Z.Y.M.\ through grant WE~3649/2-1.
\end{acknowledgments}



\bibliography{grapheneMFT-ED-QMC}


\end{document}